\documentclass[pdflatex,sn-mathphys-num]{sn-jnl}% Math and Physical Sciences Numbered Reference Style
\usepackage{graphicx}%
\graphicspath{{Figs/}}
\usepackage{multirow}%
\usepackage{amsmath,amssymb,amsfonts,upgreek}%
\usepackage{amsthm}%
\usepackage{mathrsfs}%
\usepackage[title]{appendix}%
\usepackage[dvipsnames]{xcolor}
\definecolor{HL}{named}{BrickRed}
\usepackage{textcomp}%
\usepackage{manyfoot}%
\usepackage{booktabs}%
\usepackage{bm}
\usepackage{mleftright}
\theoremstyle{thmstyleone}%
\theoremstyle{thmstyletwo}%

\theoremstyle{thmstylethree}%

\def\a{\alpha}
\def\b{\beta}
\def\d{\delta}

\def\g{\gamma}

\def\e{\epsilon}

\def\p{\pi}
\def\s{\sigma}
\def\f{\phi}
\def\m{\mu}

\def\w{\omega}
\newcommand{\bt}{\bar{t}}
\newcommand{\hd}{\hat{d}}

\def\G{\Gamma}
\def\W{\Omega}
\newcommand{\cG}{\mathcal{G}}
\newcommand{\ii}{\mathrm{i}}
\DeclareMathOperator{\re}{Re}
\DeclareMathOperator{\im}{Im}
\DeclareMathOperator{\tr}{Tr}
\newcommand{\Ar}{\mathrm{A}}
\newcommand{\Rr}{\mathrm{R}}

\begin{document}

\title[Article Title]{Nonequilibrium transport through the Hubbard dimer}

\author[1]{\fnm{Yaroslav} \sur{Pavlyukh}}\email{yaroslav.pavlyukh@pwr.edu.pl}
\author[2]{\fnm{Riku} \sur{Tuovinen}}\email{riku.m.s.tuovinen@jyu.fi}
\affil[1]{%
  \orgdiv{Institute of Theoretical Physics, Faculty of Fundamental Problems of Technology}, %
  \orgname{Wroclaw University of Science and Technology}, %
  \orgaddress{\city{Wroclaw}, \postcode{50-370}, \country{Poland}}}
\affil[2]{%
  \orgdiv{Department of Physics, Nanoscience Center}, %
  \orgname{University of Jyv{\"a}skyl{\"a}}, %
  \orgaddress{P.O. Box 35, \postcode{40014}, \country{Finland}}}

\abstract{We apply a computationally efficient approach to study the time- and energy-resolved spectral properties of a two-site Hubbard model using the nonequilibrium Green’s function formalism. By employing the iterative generalized Kadanoff-Baym ansatz ($i$GKBA) within a time-linear framework, we avoid the computational cost of solving the full two-time Kadanoff-Baym equations. Spectral information is extracted by coupling the system to multiple narrow-band leads, establishing a direct analogy to photoemission experiments. Our results reveal correlation-induced shifts and broadenings of spectral features, along with a suppression of transient current oscillations. This approach provides a promising avenue for analyzing correlated electron dynamics in open quantum systems.}




\maketitle

\section{Introduction}\label{sec:intro}
\begin{figure}[t]
  \centering
  \includegraphics[width=0.45\textwidth]{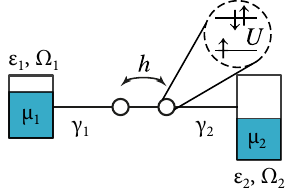}
  \hspace{0.5cm}
  \includegraphics[width=0.45\textwidth]{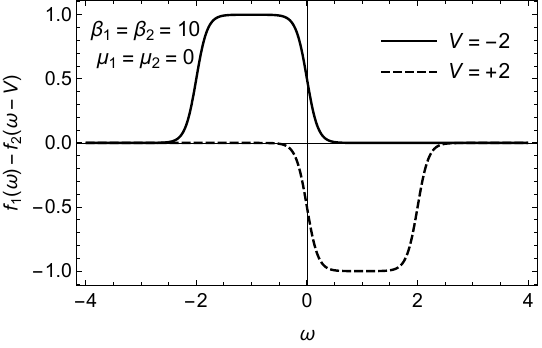}
\caption{Left: system setup consisting of a Hubbard dimer connected to two leads with different chemical potentials $\mu_\a$, energy centroids $\epsilon_\a$ and bandwidths $\W_\a$, right: difference of the distribution functions in a biased system\,---\,ingredient of the Landauer-B\"{u}ttiker formula (Eq.~\ref{eq:LB}) for steady-state currents.}\label{fig:1}
\end{figure}

Hubbard model~\cite{hubbard_electron_1963} is one of the most studied models of strongly correlated systems. It is traditionally used as a testbed of new approximations and numerical methods. As an open quantum system it was used in demonstrating heat transfer mechanisms~\cite{esslinger_fermi-hubbard_2010, bertini_finite-temperature_2021} and in studies of thermoelectricity~\cite{brantut_thermoelectric_2013,karrasch_thermal_2016}. The focus of this work is the so-called Hubbard dimer, i.e., a model with two sites in contact with electronic reservoirs (Fig.~\ref{fig:1}, left).

In our previous work~\cite{pavlyukh_nonequilibrium_2024}, based on the nonequilibrium Green’s function (NEGF) approach, we demonstrated that coherent electron dynamics of the isolated Hubbard dimer can be studied analytically in four important approximations: second Born, $GW$, and two flavors of $T$-matrix approximation. It was shown that the off-diagonal elements of the density matrix satisfy oscillator-like equations of motion. This enabled us to derive analytical expansions for the system’s equilibrium properties, as reached through an adiabatic switching protocol starting from a non-interacting initial state. This approach serves as an alternative to equilibrium many-body perturbation theory (MBPT)~\cite{romaniello_self-energy_2009,carrascal_hubbard_2015,di_sabatino_scrutinizing_2021}.

A crucial ingredient of these derivations is the generalized Kadanoff–Baym ansatz (GKBA)~\cite{lipavsky_generalized_1986} which reduces the complex integro-differential Kadanoff–Baym equations (KBE) to a set of coupled ordinary differential equations (ODE)~\cite{schlunzen_achieving_2020,joost_g1-g2_2020,pavlyukh_photoinduced_2021}. The strength of the GKBA+ODE formulation lies in its linear scaling with physical propagation time. This significantly pushes the limits of NEGF approach as compared to solving integro-differential KB equations, which scale cubically with physical time. However, GKBA represents the lesser/greater Green's functions (GFs) in a simplified form,
\begin{align}
  G^{\lessgtr}(t,t')&=-G^{R}(t,t')\rho^{\lessgtr}(t')+\rho^{\lessgtr}(t)G^{A}(t,t'),\label{eq:e:gkba}
\end{align}
typically using a mean-field retarded propagator (specified below):
\begin{align}
G^{R}(t,t')&=-i\theta(t-t')T\left\{e^{-\ii \int_{t'}^t d\tau\, h_\text{HF}(\tau)}\right\}.\label{eq:gr:hf}
\end{align}
This makes it impossible to retrieve the spectral information contained in the full two-times GFs, $G^{\lessgtr}(t,t')$. The limitation is significant because many experimental techniques rely on spectral information to probe electronic properties.

Multidimensional coherent spectroscopies \cite{cundiff_optical_2013, cheng_dynamics_2009, huang_quantum_2023} provide a direct way to access two-time correlation functions by measuring system responses to sequences of ultrafast optical pulses. These techniques are particularly useful for studying coherent electronic dynamics but require sophisticated setups capable of resolving temporal correlations explicitly.

In contrast, photoemission spectroscopies \cite{schuurman_time-resolved_2022, huang_high-resolution_2022} do not measure two-time correlation functions directly. Instead, they provide access to electronic spectral properties by detecting emitted electrons and recording their energies. The measured signal reflects the density of states, integrated over the measurement time window, rather than explicit time-dependent correlations. Since theories based on single-time correlators cannot naturally recover two-time spectral information, modeling photoemission within such frameworks requires an alternative approach\,---\,namely, the introduction of an energy-selective detection mechanism.

To achieve this, we model detectors as an \emph{embedding} characterized by the tunneling matrix elements $T_{ik\a}$ between the system state $i$ and an environmental state  $k$ in electronic lead $\a$ with energy $E_{k\a}$. Since perfectly energy-selective detectors are unphysical, we describe the associated tunneling rate using a peaked function (Lorentzian distribution):
\begin{equation}\label{eq:lorentzian}
\Gamma_{\a,ij}(\w) = \sum_{k} T_{ik\a}T_{k\a j}\delta(E_{k\a}-\w)=\frac{\gamma_{\a,ij}\Omega_\a^2}{(\w-\e_\a)^2+\Omega_\a^2},
\end{equation}
which is centered at $\epsilon_\a$ with width $\W_\a$. Similar concepts have been employed previously, for example, in Ref.~\cite{schuler_time-dependent_2016}, where a continuum embedding was used to describe photoemission from a quasiparticle peak and its plasmon satellite. However, extending this approach to capture transient dynamics across a broad energy range requires an efficient method for propagating the system’s state in time. With the advent of time-linear GKBA-based methods, such modeling becomes computationally feasible. To illustrate the potential of this approach, we apply it to study bias-driven transient dynamics in a Hubbard dimer.

The outline of this work is as follows: In Sec.~\ref{sec:theory}, we introduce the GKBA+ODE approach. We begin with a description of correlated electron dynamics in closed systems and then generalize it to open systems in contact with a structured environment. Next, we review the electronic properties of the two-site Hubbard model. In Sec.~\ref{sec:results}, we present our results, focusing on time- and energy-resolved currents. Finally, in Sec.~\ref{sec:conclusion}, we summarize our findings.
\section{Theory}\label{sec:theory}
\subsection{Time-linear nonequilibrium Green's function formalism}
In the NEGF formalism, the electronic lesser/ greater single-particle Green’s functions
\begin{align}
  G^{<}_{ij}(t,t')&= i\langle \hat{d}_j^\dagger(t')\hat{d}_i(t)\rangle,&
  G^{>}_{ij}(t,t')&=-i\langle \hat{d}_i(t)\hat{d}_j^\dagger(t')\rangle,
  \label{eq:def:G}
\end{align}
are the fundamental correlators giving rise to various physical observables. The annihilation and creation operators $\hat{d}^{(\dagger)}$ will be specified later with the model description. In time-linear ODE formulations, instead of solving the full two-times Kadanoff-Baym equations:
\begin{align}
\left[ i \partial_{t} - h(t) \right] G^\lessgtr(t,t')
=\left [\Sigma^\lessgtr \cdot G^A + \Sigma^R \cdot
G^\lessgtr \right]\!(t,t'),\label{eq:e:KBE}
\end{align}
where $\left[ A \cdot B\right](t,t') \equiv \int d\bt\,A(t,\bt) B(\bt,t'),$ is a real-time convolution, 
$X^{R/A}(t,t')$ are the retarded/advanced functions, and $\Sigma$ is the correlation part of the self-energy,
one aims to find a system of equations for the time-diagonal of GFs, which are the densities:
\begin{align}
  \rho_{ij}^{\lessgtr}(t)=-i G^{\lessgtr}_{ij}(t,t).
\end{align}
Combining Eq.~\eqref{eq:e:KBE} with its adjoint and going to the equal times limit one
obtains:
\begin{align}
  \frac{d}{dt}\rho^<(t)&=-i\big[h_\text{HF}(t),\rho^<(t) \big] -\left(I(t)+I^\dagger(t)\right),
  \label{eq:eomrho:e}
\end{align}
where $h_{\text{HF}}(t)=h(t)+V_{\text{HF}}(t)$ is the Hartree-Fock Hamiltonian. The mean-field potential is given by
\begin{align}
  V_{\text{HF},ij}(t)&=\sum_{mn}w_{imnj}\rho^{<}_{nm}(t),
\end{align}
with $w_{imnj}=v_{imnj}-v_{imjn}$,  and $v_{imnj}$ being the Coulomb matrix elements. The collision term is expressed in terms of the two-particle Green's function (2-GF)
\begin{align}
  I_{lj}(t)&=I^\text{c}_{lj}(t)=-i\sum_{imn} v_{lnmi}(t) \cG_{imjn}(t),\label{eq:i:ee}
\end{align}
which is a functional of the two-times $G^\lessgtr(t,t')$ and which can be expressed in terms of $\rho_{ij}^{\lessgtr}(t)$ with the help of GKBA using an appropriate approximation for the correlated self-energy. The procedure is described in detail in Ref.~\cite{pavlyukh_time-linear_2022-1}, and its application to the Hubbard dimer is given in Ref.~\cite{pavlyukh_nonequilibrium_2024}. In the present work we focus on the so-called second Born approximation for the electron self-energy. This approximation works very well for the dimer at half-filling as demonstrated for spectral
properties~\cite{romaniello_beyond_2012} and densities~\cite{pavlyukh_nonequilibrium_2024}. Extensive comparison of the self-energy approximations out of equilibrium was performed by Schl\"{u}nzen \emph{et al.}~\cite{schlunzen_nonequilibrium_2017,schlunzen_ultrafast_2020}.

Besides electronic correlations, the GKBA+ODE approach can also incorporate other physical processes, such as interaction with bosonic particles~\cite{karlsson_fast_2021} and environment~\cite{tuovinen_time-linear_2023}. In this case, the integral $I(t)$ becomes a sum of different collision mechanisms. It was shown that interaction with cavity photons gives rise to the Autler-Townes spectral features~\cite{pavlyukh_interacting_2022} and is manifested in the high harmonics spectrum excited in a transport setup~\cite{tuovinen_electroluminescence_2024}. In this work, we focus on the interplay of electronic correlations and transport, therefore the collision integral consists of the correlation (c) and embedding (em) terms
\begin{align}
  I(t)=I^{\text{c}}(t)+I^{\text{em}}(t).
\end{align}

For wide-band leads, i.e., $\W_\a\rightarrow\infty$, the embedding self-energy takes a simple form~\cite{stefanucci_nonequilibrium_2013}
\begin{subequations}
\begin{align}
	\Sigma^{R}_{\a}(t,t')&=-\frac{i}{2}s^{2}_{\a}(t)\d(t,t')\,\g_{\a},
	\\
	\Sigma^{<}_{\a}(t,t')&=is_{\a}(t)s_{\a}(t')e^{-i\f_{\a}(t,t')}\!\!
	\int\!\!\frac{d\w}{2\p}f_\a(\w)e^{-i\w(t-t')}\,\g_{\a},
\end{align}
\end{subequations}
where $s_{\a}(t)$ is the switch-on function for the contact  between the system and electrode $\a$, $\f_{\a}(t,t')\equiv \int_{t'}^{t}d\bar{t}\;V_{\a}(\bar{t})$ is the accumulated phase  due to the time-dependent voltage $V_{\a}$, and $f_\a(\w)=1/(e^{\upbeta_\a(\w-\m_\a)}+1)$ is the Fermi function at inverse  temperature $\upbeta_\a$ and chemical potential $\m_\a$.
Expanding into the partial fractions~\cite{hu_communication:_2010}
\begin{multline}
  f_\a(\omega)=\frac12-\sum_{\ell\ge 1}\eta_\ell
  \Bigl[\frac{1}{\upbeta_\a(\w-\mu_\a)+i\zeta_\ell}+\frac{1}{\upbeta_\a(\w-\mu_\a)-i\zeta_\ell}\Bigr],\quad\text{with\; $\re \zeta_\ell>0$,}\label{eq:pole:exp}
\end{multline}
allows to express $I^{\text{em}}(t)$ in terms of an embedding correlator $\cG^{\text{em}}(t)$~\cite{tuovinen_time-linear_2023}
\begin{align}
  I^{\text{em}}(t)&=-\frac{1}{4}\G(t)-\sum_{\ell\a} s_{\a}(t)\frac{\eta_\ell}{\upbeta_\a}\G_{\a}\cG^{\text{em}}_{\ell\a}(t)
\end{align}
that fulfills the equation of motion:
\begin{align}
\!\! i\frac{d}{d t} \cG^{\text{em}}_{\ell\a}(t)  = - s_{\a}(t) - 
 \cG^{\text{em}}_{\ell\a}(t)\Big(h^{\dag}_{\text{eff}}(t)-V_\a(t)-\mu_\a+i\frac{\zeta_\ell}{\upbeta_\a}\Big).
\label{eomgla}
\end{align}
Here, $h_{\text{eff}}(t)\equiv h_{\text{HF}}(t)-i\G(t)/2$ is the effective mean-field Hamiltonian and $\G(t)=\sum_{\a}s^{2}_{\a}(t)\g_{\a}$. This approach was recently generalized~\cite{pavlyukh_open_2025} towards finite widths spectral densities (Eq.~\ref{eq:lorentzian}) incorporating reconstructions~\cite{kalvova_beyond_2019, kalvova_dynamical_2023, kalvova_fast_2024, kalvova_short-time_2025} of $G^{\lessgtr}(t,t')$ on the level beyond GKBA (Eq.~\ref{eq:e:gkba}). The generalization, referred to as the \emph{iterated generalized Kadanoff-Baym ansatz} ($i$GKBA), follows the same basic principles, albeit at the cost of introducing additional correlators. These correlators must be co-propagated alongside the equation of motion for the density~\eqref{eq:eomrho:e}, contributing to the electron dynamics and enabling the determination of the time-resolved electronic currents through a generalization of the Meir, Wingreen and Jauho~\cite{meir_landauer_1992,jauho_time-dependent_1994} formula.

The computational complexity of propagating $\cG_{imjn}(t)$ and computing the electronic collision integral $I^{\text{c}}(t)$ is $\mathcal{O}(N_{\text{$e$-basis}}^5N_t)$, where $N_t$ is the number of time-steps. The computational complexity of the embedding collision integral $I^{\text{em}}(t)$ at the $i$GKBA level is  $\mathcal{O}(N_{\text{$e$-basis}}^3N_{\text{leads}}^2(2N_{\text{leads}}+4N_p)N_t)$, where $N_p$ is the number of poles in the expansion~\eqref{eq:pole:exp} of the Fermi-Dirac distribution function. This is comparable with the cost of solving the Dyson equation for the equilibrium Green's function. This approach typically consists of evaluating the second Born self-energy and solving the Dyson equation until self-consistency is achieved. The self-energy is typically evaluated in the imaginary time-domain on a grid of $N_\tau$ points with $\mathcal{O}( N_{\text{$e$-basis}}^5N_\tau)$ complexity~\cite{schuler_spectral_2018}. The scaling of solving the Dyson equation strongly depends on the domain and representation, as reviewed in Ref.~\cite{dong_legendre-spectral_2020}. The analytic continuation of the Green's function from imaginary to real frequency, which is an important ingredient of such an approach, is a separate and intensely investigated issue~\cite{fei_nevanlinna_2021}.

\subsection{System}
Before applying the $i$GKBA method to the Hubbard dimer, let us recall its relevant electronic properties. Its Hamiltonian in the site-spin basis $i\equiv(\bm{i},\sigma_i)$ is given by:
\begin{equation}
\hat H=h \sum_\sigma\sum_{\bm{i}\neq\bm{j}} \hd_{\bm{i}\sigma}^\dagger \hd_{\bm{j}\sigma}
+U\sum_{\bm{i}}\hat{n}_{\bm{i}\uparrow}\hat{n}_{\bm{i}\downarrow}
-\mu\sum_{\bm{i}\sigma}\hat{n}_{\bm{i}\sigma},
\end{equation}
where $h$ is the hopping parameter (subsequently set to 1), $U$ is the on-site repulsion and $\hat{n}_{\bm{i}\sigma}=\hd_{\bm{i}\sigma}^\dagger \hd_{\bm{i}\sigma}$. The chemical potential is fixed at $\mu=\frac{U}{2}$. The density matrix at half-filling is specified by a single parameter $a$ and reads in the site basis:
\begin{equation}
  \rho_\s^<=\begin{pmatrix}
  \frac12&a\\
  a&\frac12
\end{pmatrix}.\label{eq:rho}
\end{equation}
The Hartree-Fock Hamiltonian is independent of $a$ and is identical to the hopping part of the Hamiltonian:
\begin{equation}
h_\text{HF}=\begin{pmatrix}
0&h\\
h&0
\end{pmatrix},\label{eq:h:hf}
\end{equation}
with eigenvalues $\epsilon_i^\text{HF}=\pm h$.

At equilibrium, the isolated system has been studied using a large number of approximations. Its spectral density (see Fig.~1 of Ref.~\cite{di_sabatino_scrutinizing_2021}, Fig.~6 of Ref.~\cite{romaniello_beyond_2012} or Fig.~13 of Ref.~\cite{carrascal_hubbard_2015}) consists of two quasiparticle peaks with energies $\epsilon_i \stackrel{U\rightarrow0}{\rightarrow} \epsilon_i^\text{HF}$, each accompanied by a satellite peak. In the atomic limit, i.e., as $h/U\rightarrow 0$, the spectral weight is equally distributed between the peaks. We probe the system by connecting it to two leads and applying the bias voltage $V_2(t)$ to the second lead. For negative or positive biases, occupied or unoccupied states are probed, respectively (Fig.~\ref{fig:1}, right).

Under equilibrium conditions the electronic density is stationary and, in general, deviates from the form~\eqref{eq:rho} due to the tunneling to the leads in the presence of electronic correlations. The simplest, albeit numerically demanding, way to obtain the equilibrium density is by applying the adiabatic switching procedure, in which the electron interaction and the tunneling matrices become time-dependent, i.e., $U(t)=s_0(t)U$, $T_{ik\a}(t)=s_\a(t)T_{ik\a}$, typically we use the ramp functions of the form  $s_{\a}(t)=\cos\mleft(\pi/2\cdot t/t_i\mright)^2\theta(-t)+\theta(t)$, which build up correlations in the system over the time interval $[t_i,0]$.

\subsection{Time-dependent and steady-state currents}
Since in GKBA formalism the collision integrals are known, time-dependent currents can be determined directly from the Meir-Wingreen formula~\cite{meir_landauer_1992,jauho_time-dependent_1994}
\begin{align}
  J_\a(t)&=2\re\tr\mleft[I_\a(t)\mright],
\end{align}
where $I^\text{em}(t)=\sum_\a I_\a(t)$, and individual contributions of each lead $I_\a(t)$ are determined from the $\cG_{\ell\a}^\text{em}(t)$ correlator~\eqref{eomgla} in the case of GKBA, and from several correlators in the case of $i$GKBA~\cite{pavlyukh_open_2025}.

Under stationary conditions, the steady-state currents in noninteracting systems or systems treated at the mean-field level can be determined from the Landauer-B\"{u}ttiker (LB) formula~\cite{stefanucci_nonequilibrium_2013}
\begin{multline}
  J_{\a}=\int\frac{d\w}{2\pi}\sum_\b\left(f_\a(\w-V_\a)-f_\b(\w-V_\b)\right)\\
  \times\tr\mleft[
    \G_\a(\w-V_\a) G_0^{\Rr}(\w)\G_\b(\w-V_\b) G_0^{\Ar}(\w)
    \mright].\label{eq:LB}
\end{multline}
Here, $f_\a$ and $f_\b$ are the distribution functions associated with leads $\a$ and $\b$, respectively, and $G_0^{\Ar/\Rr}$ are the equilibrium mean-field advanced/retarded Green's functions of the central system. Extensions of this result to transient scenarios have also been demonstrated within the wide-band limit approximation~\cite{tuovinen_time-dependent_2013,tuovinen_time-dependent_2014} with further generalizations~\cite{ridley_current_2015,ridley_fluctuating-bias_2016,ridley_many-body_2022,ridley_photon-assisted_2025}. 

Once the density matrix is known, mean-field currents can be easily computed using Eq.~\eqref{eq:LB}. This follows from the fact that, at the Hartree-Fock level, the electron self-energy consists only of the embedding part, $\Sigma(\w)=\Sigma_{\text{em}}(\w)=\sum_\a\Sigma_\a(\w)$, where 
\begin{align}
  \im  \Sigma_\a^{\Rr}(\w)&=-\frac12\G_\a(\w),&
  \re  \Sigma_\a^{\Rr}(\w)&=\frac12 \mathcal{H}[\G_\a](\w).
\end{align}
Here, $\mathcal{H}$ is the Hilbert transform:
\begin{align}
  \mathcal{H}[x](\w)&=\frac{1}{\p}\mathcal{P}\int_{-\infty}^{\infty}\!d\nu\,\frac{x(\nu)}{\w-\nu}.
\end{align}
We have then
\begin{align}
  G^{\Rr}_{0}(\w)&=\frac{1}{\w-h_{\text{HF}}-\Sigma^{\Rr}(\w)},
\end{align}
and $G^{\Ar}_{0}(\w)=[G^{\Rr}_{0}(\w)]^\dagger$.

Despite being applicable only to noninteracting system, Eq.~\eqref{eq:LB} represents a useful starting point for understanding more complicated interacting and time-dependent scenarios. On its basis it can be concluded that the current is non-zero if there is a non-zero overlap between $f_\a-f_\b$, the spectral densities $\G_\a$, $\G_b$ and the density of states of the central system. By varying the energy position $\e_\a$ of a narrow-band left lead ($\a=1$), and using a wide-band $\W_\b\rightarrow \infty$ right lead ($\b=2$), the desired energy selectivity can be achieved.

It should be noted that Eq.~\eqref{eq:LB} is not the only way to understand spectral properties. In a series of papers G. Cohen \emph{et al.}~\cite{cohen_greens_2014, cohen_greens_2014-1}  gave a prescription how the spectral function of a system can be determined starting the Meir-Wingreen expression for the steady-state current~\cite{stefanucci_nonequilibrium_2013}:
\begin{align}
J_\a  = -2i \int \frac{d\w}{2\pi} \tr\mleft[ \Sigma^<_\a(\w)A(\w)-\Gamma_\a(\w)G^<(\w)
  \mright],\label{eq:MW}
\end{align}
where the lesser component of the lead specific embedding self-energy can be expressed in terms of the distribution function $f_\a(\w)$ and the tunneling rate matrix $\Gamma_\a(\w)$ as $\Sigma^<_\a(\w)=if_\a(\w)\Gamma_\a(\w)$, but $G^<(\w)$ is not known within GKBA.  Unlike the Landauer–Büttiker formula, this approach does not require the assumption of a noninteracting system: the unknown lesser Green's function is eliminated by taking the difference between currents computed using fully occupied and completely empty auxiliary leads at varying energy positions. This technique\,---\,varying the energy positions of two auxiliary leads\,---\,is complementary to the approach of varying a single lead's chemical potential. The latter idea is more similar to the surface tunneling spectroscopy setup and in the context of GKBA was discussed in Ref.~\cite{cosco_interacting_2024}.

The method of G. Cohen \emph{et al.}~\cite{cohen_greens_2014, cohen_greens_2014-1}, involving two auxiliary leads, provides a rigorous means of extracting the spectral function from measured currents. However, this scheme cannot be achieved in photoemission spectroscopies, as the occupation of detectors cannot be externally controlled. Therefore, in the present work, we consider a scenario involving multiple narrow-band leads at varying energy positions but fixed chemical potential (e.g., aligned with that of the central system). This setup enables exploration of the entire spectral range in a single simulation. We note that the two-lead auxiliary approach is fully compatible with the $i$GKBA formalism and will be employed in forthcoming systematic studies of electron correlation effects.

\section{Results}\label{sec:results}

\begin{figure}[t]
\centering
\includegraphics[width=0.99\textwidth]{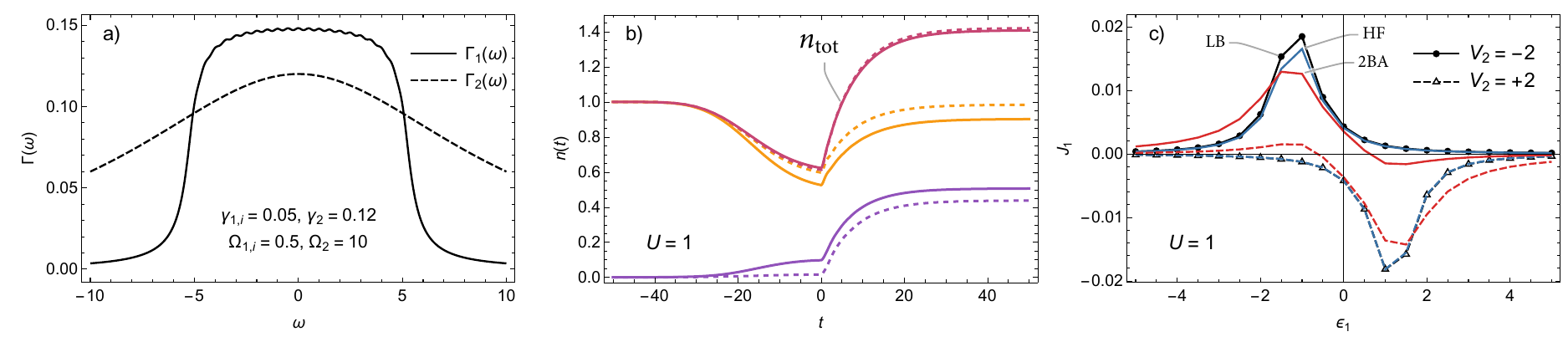}
\includegraphics[width=0.4\textwidth]{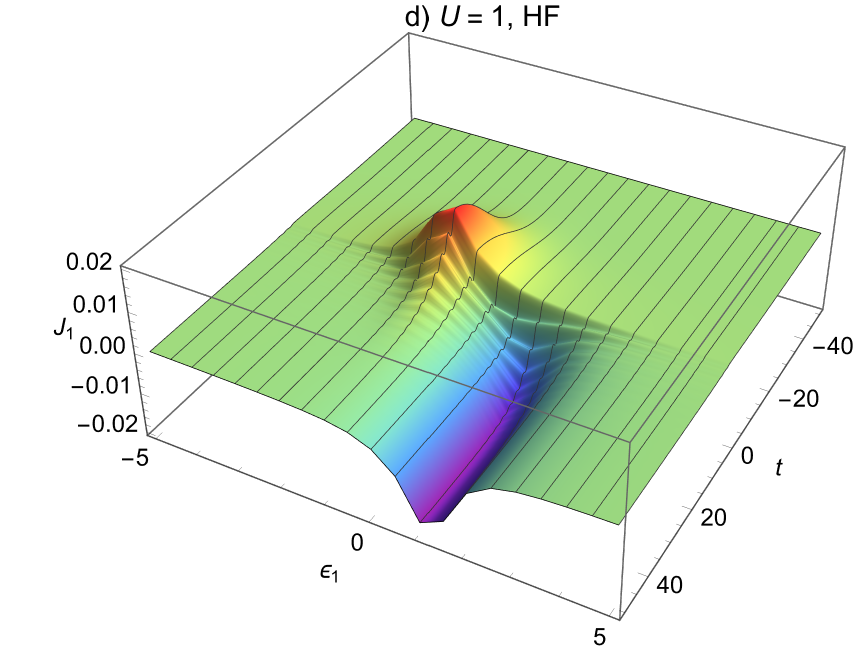}
\includegraphics[width=0.4\textwidth]{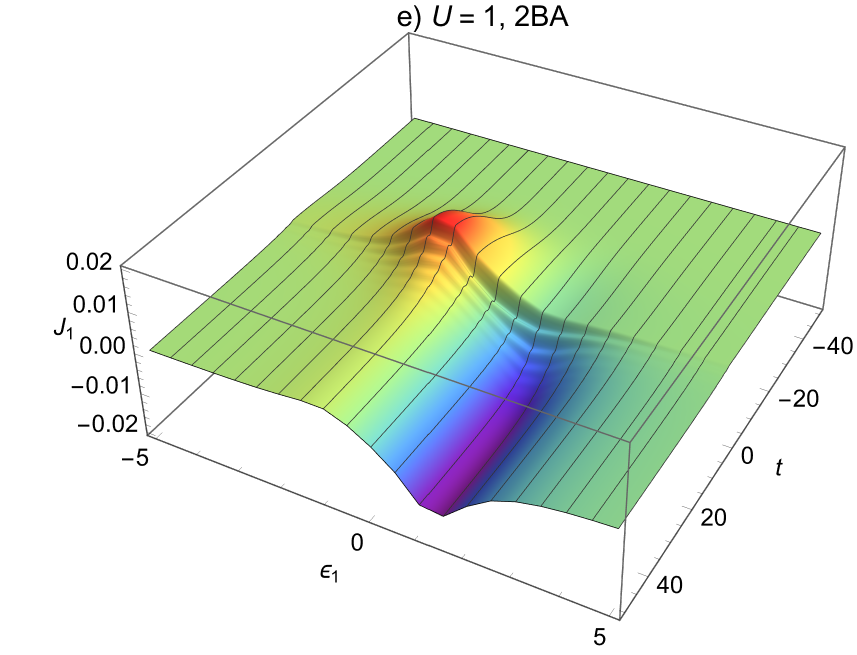}
\caption{
Weakly correlated Hubbard dimer with $U=1$. Top: a) spectral function of the leads, b) evolution of natural occupations (purple and orange), total number of electrons in the central system, dashed lines\,---\,Hartree-Fock, solid lines\,---\,second Born approximation, and c) energy-dependent currents through the left lead at stationary states corresponding to two bias voltages $V_2$ applied to the right lead; The currents are computed for predetermined energies $\e_{1,j}$ of the subleads, continuous lines are a guide to the eye. Bottom: time- and energy-resolved currents through the left lead using d) the Hartree-Fock and e) second Born approximations.}\label{fig:U:1}
\end{figure}

We start with a weakly correlated case of $U=1$, Fig.~\ref{fig:U:1}. The left lead is composed of 21 subleads of spectral widths $\W_{1,j}=0.5$, centered at $\e_{1,j}=-5+0.5(j-1)$ for $1\le j\le 21$. The right lead is centered at $\e_2=0$, and its width is $\W_2=10$. The leads' inverse temperatures for all calculations are fixed at $\upbeta_{1,j}=\upbeta_2=10$. The tunneling matrices in site basis are given by $\gamma_{1,j}=0.05\begin{pmatrix}1&0\\0&0\end{pmatrix}$ and $\gamma_{2}=0.07\begin{pmatrix}0&0\\0&1\end{pmatrix}$. In panel (a), the composite spectral function (its trace) of the left lead [$\G_1(\w)$] is compared with the spectral function of the right lead [$\G_2(\w)$].  The system is driven by the bias voltage $V_2(t)$ according to the following protocol:
\begin{align}
  V_2(t)&=-2+4[1+\exp(-25t)]^{-1}.
\end{align}
This means that during the adiabatic switching the system is subject to $V_2(t)=-2$. Then, at $t\sim0$, the bias very quickly (on the order of 0.04 time units) changes its sign and the system evolves under $V_2(t)=2$. In panel (b), the mean-field (HF, dashed lines) and the second Born approximation (2BA, solid lines) total ($\tr\mleft[\rho_\sigma^<\mright]$) and  natural (eigenvalues of the density matrix $\rho_\sigma^<$) occupations are compared. During the first time-interval the system loses approximately 0.4 electrons in each spin channel and during the second time-interval the system reaches $1.4$ total number of electrons. Both approximations are consistent in the value of total occupation $n_{\text{tot}}(t)$, but they differ with respect to the natural occupations.

In Fig.~\ref{fig:U:1}(c), currents at $t=-1$ and $t=t_f$ (end of propagation) are compared. The LB currents are computed using the HF electron density matrix, they are in a good agreement with the HF currents obtained by the time propagation. Small deviations for $V_2=-2$ are explained by incomplete thermalisation during the first time interval. For the second time-interval, the system thermalizes faster because it starts with already a correlated state from the first time-interval. 2BA results (red lines) closely agree with the HF results (blue lines), whereas peaks are slightly broader. Energy- and time-resolved currents are compared in panels (d) and (e), showing similarity between HF and 2BA in the weakly correlated case.

\begin{figure}[t]
\centering
\includegraphics[width=0.99\textwidth]{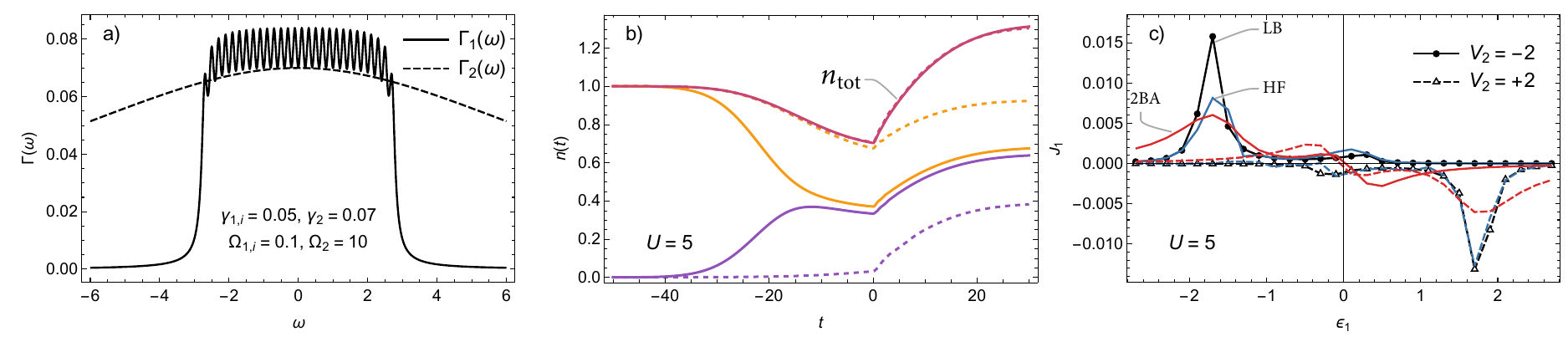}
\includegraphics[width=0.4\textwidth]{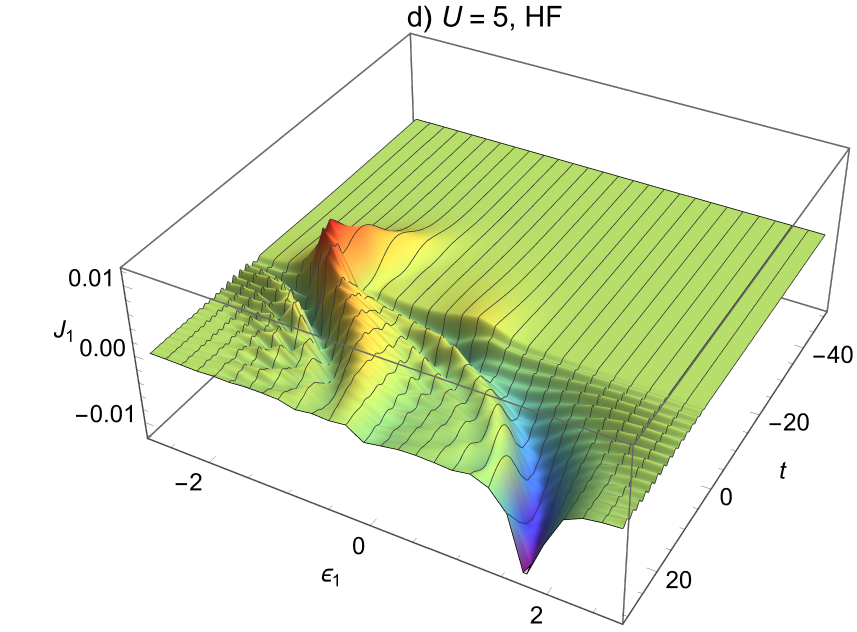}
\includegraphics[width=0.4\textwidth]{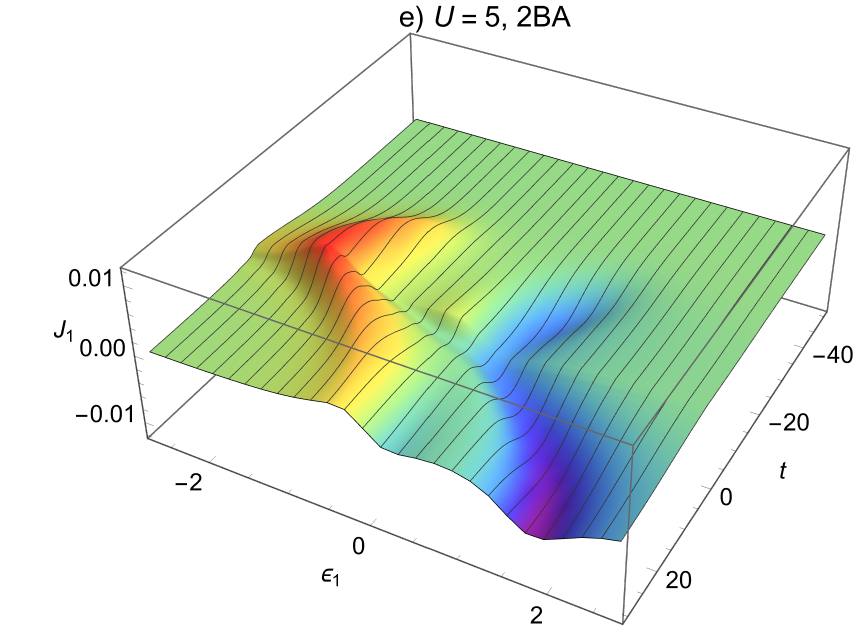}
\caption{Moderately correlated Hubbard dimer with $U=5$. Top: a) spectral function of the leads, b) evolution of natural occupations (purple and orange), total number of electrons in the central system, dashed lines\,---\,Hartree-Fock, solid lines\,---\,second Born approximation, and c) energy-dependent currents through the left lead at stationary states corresponding to two bias voltages $V_2$ applied to the right lead; The currents are computed for predetermined energies $\e_{1,j}$ of the subleads, continuous lines are a guide to the eye. Bottom: time- and energy-resolved currents through the left lead using d) the Hartree-Fock and e) second Born approximations.}\label{fig:U:5}
\end{figure}

\begin{figure}[t]
\centering
\includegraphics[width=0.99\textwidth]{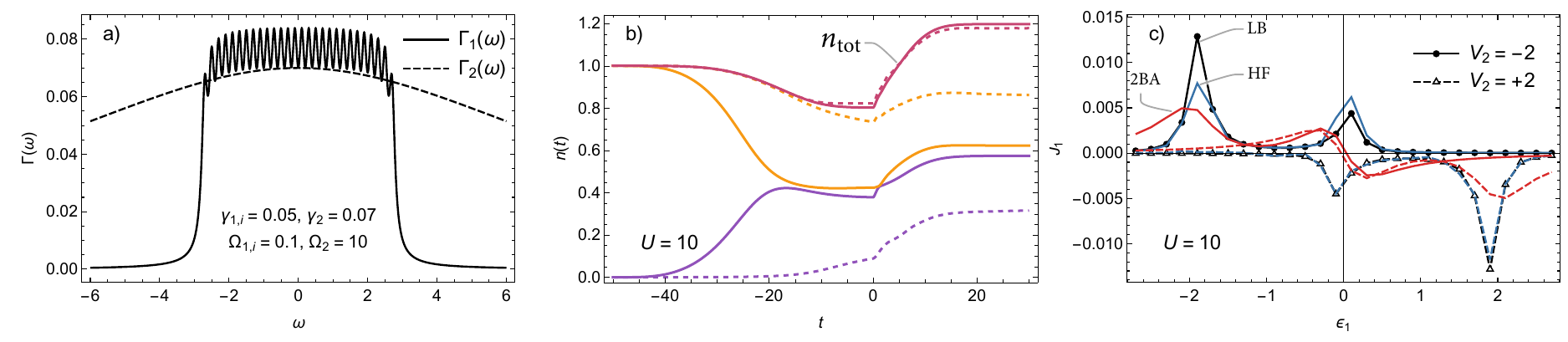}
\includegraphics[width=0.4\textwidth]{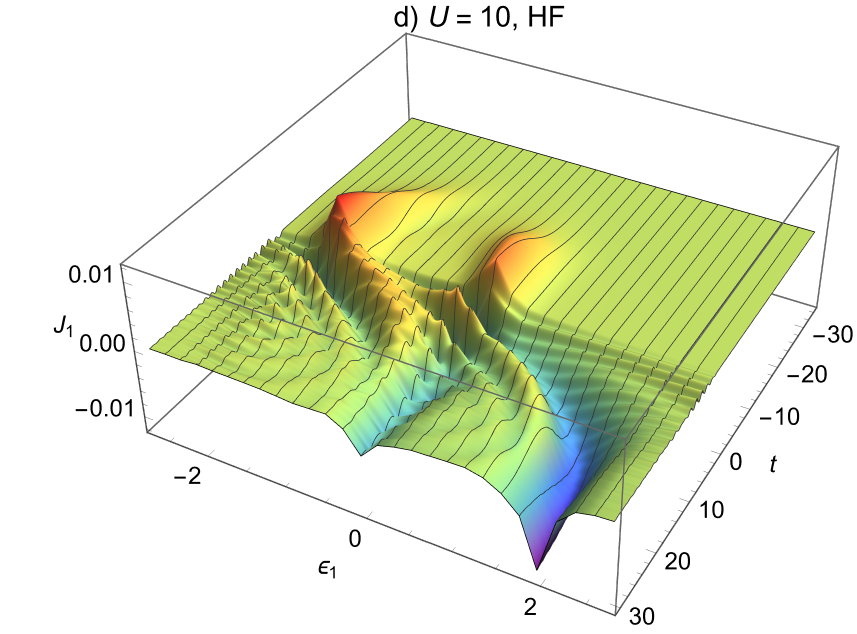}
\includegraphics[width=0.4\textwidth]{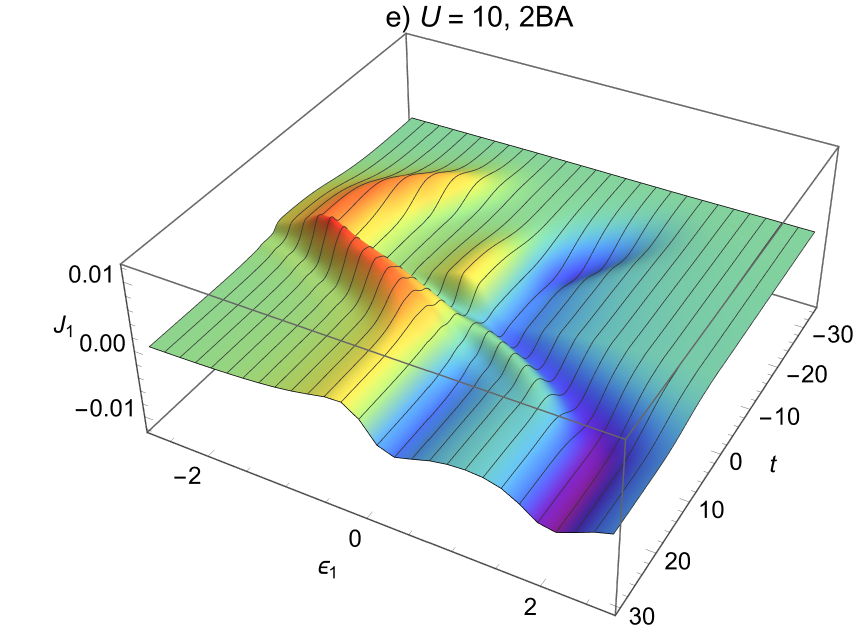}
\caption{Strongly correlated Hubbard dimer with $U=10$. Top: a) spectral function of the leads, b) evolution of natural occupations (purple and orange), total number of electrons in the central system, dashed lines\,---\,Hartree-Fock, solid lines\,---\,second Born approximation, and c) energy-dependent currents through the left lead at stationary states corresponding to two bias voltages $V_2$ applied to the right lead; The currents are computed for predetermined energies $\e_{1,j}$ of the subleads, continuous lines are a guide to the eye. Bottom: time- and energy-resolved currents through the left lead using d) the Hartree-Fock and e) second Born approximations.}\label{fig:U:10}
\end{figure}

For calculations in the moderately- ($U=5$) and strongly-correlated ($U=10$) regimes, Figs.~\ref{fig:U:5} and \ref{fig:U:10}, slightly different lead parameters are selected: in order to increase the energy resolution we reduce the coupling $\gamma_{2}=0.07$ and the width of subleads $\W_{1,j}=0.1$. However, to be able to cover the energy interval from $-3$ to $3$, the number of subleads is increased to 28, such that they are positioned at $\epsilon_{1,j}=-2.7+0.2(j-1)$.

There are three pronounced effects of electronic correlations: i) A substantial broadening of the current peaks (panel c). ii) As a consequence, there is a suppression of the transient current oscillations (panel e), which are highly pronounced in the Hartree-Fock case (panel d). As explained in Ref.~\cite{stefanucci_nonequilibrium_2013, tuovinen_time-dependent_2013,tuovinen_time-dependent_2014}, these oscillations originate from virtual transitions between the resonant levels of the central system and the Fermi level of the biased reservoirs. Full Kadanoff-Baym calculations for isolated Hubbard clusters similarly demonstrate a suppression of the density oscillations at the fully self-consistent 2BA level (Fig.~4(d) of Ref.\cite{von_friesen_successes_2009}). iii) Finally, the position of the peaks varies significantly during the time evolution. This aspect can be attributed to the large contribution of the mean-field potential at large  $U$-values and a substantial evolution of the total electron density (panel b). Remarkably, the total electron occupation of the Hubbard dimer is not sensitive to the choice of method, as also observed in Ref.~\cite{uimonen_comparative_2011, khosravi_correlation_2012}.

The contribution of satellite peaks to the electron current is a separate and complicated issue. We were unable to resolve them at $U=1$ because, in this case, the spectral weight of the satellites is small. For larger $U$-values, the satellites approach the main quasiparticle peak, requiring higher energy resolution. A good reference point would be the KBE calculations of Ref.~\cite{von_friesen_successes_2009, puig_von_friesen_kadanoff-baym_2010}. Comprehensive identification of all correlation features is a major undertaking and will be addressed in forthcoming publications.

\section{Conclusions}\label{sec:conclusion}
In this work, we demonstrate the feasibility of accessing the time- and energy-resolved spectral density of a two-site Hubbard model using a nonequilibrium Green's function formalism. Our approach avoids the computationally demanding solution of the full two-time Kadanoff-Baym equation. Instead, we rely on a time-linear ODE approach for open-quantum systems based on the iterative GKBA reconstruction ($i$GKBA)~\cite{pavlyukh_open_2025}. The spectral density is accessed by bringing the system into contact with two leads and computing the electron current through the narrow-band leads. This setup directly mimics a photoemission experiment. Under stationary conditions, the results of the simulations are interpreted using the Landauer-Büttiker formula. We demonstrate electronic correlations-induced shifts of energy levels and their broadening, as well as the suppression of oscillations in transient currents\,---\,effects known to affect equilibrium transport~\cite{strange_self-consistent_2011} or observed in time-dependent quantum transport through the use of computationally extensive embedded Kadanoff-Baym method~\cite{myohanen_image_2012}.
\backmatter
\bmhead{Acknowledgements}
R.T. acknowledges the financial support of the Research Council of Finland through the Finnish Quantum Flagship (Project No. 359240) and the Jane and Aatos Erkko Foundation (Project EffQSim).
\section*{Data availability}
Manuscript has no associated data.
%\bibliography{MyLibrary}

\end{document}